# Density Functional Theory of Material Design: Fundamentals and Applications – II

Ashish Kumar[1], Prashant Singh[2] and Manoj K. Harbola[3,*]

***Corresponding author email**: mkh@iitk.ac.in

[1]School of Physical Sciences, National Institute of Science Education and Research, Bhubaneswar – 752052, India
[2]Ames Laboratory, U.S. Department of Energy, Iowa State University, Ames, IA 50011, USA
[3]Department of Physics, Indian Institute of Technology, Kanpur – 208016, India

## Abstract

This is the second and the final part of the review on density functional theory (DFT), referred to as DFT-II. In the first review, DFT-I, we have discussed wavefunction-based methods, their complexity, and the basic of density functional theory. In DFT-II, we focus on fundamentals of DFT and their implications for the betterment of the theory. We start our presentation with the exact DFT result followed by the concept of exchange-correlation (xc) or Fermi-Coulomb hole and its relation with xc energy functional. We also provide the exact conditions for the xc-hole, xc-energy and xc-potential along with their physical interpretation. Next, we describe the extension of DFT for non-integer numbers of electrons, the piecewise linearity of total energy and discontinuity of chemical potential at integer particle numbers, and derivative discontinuity of the xc potential, which has consequences on fundamental gap of solids. After that, we present how one obtain more accurate xc energy functionals by going beyond LDA. We discuss the gradient expansion approximation (GEA), generalized gradient approximation (GGA), and hybrid functional approaches to designing better xc energy functionals that give accurate total energies but fail to predict properties like the ionization potential and the band gap. Thus, we describe different methods of modelling these potentials and the results of their application for the calculation of the band gaps of different solids to highlight accuracy of different xc potential. Finally, we conclude with a glimpse on orbital-free density functional theory and the machine learning approach .

In the previous article [1], referred to as DFT-I in this paper, we had discussed solution of the many-electron Schrödinger equation in detail and developed basics of density-functional theory (DFT). We ended the previous that article with a full presentation of the how the Kohn-Sham (KS) formulation of DFT is applied to a variety of systems using the local-density approximation (LDA). In the present article (DFT-II), we move to the next level of the theory and discuss many of its fundamental aspects that provide deeper insights into its working. These insights help in making implementation of DFT more accurate by facilitation development of functionals beyond the LDA. In addition, we also present calculation schemes that employ model potentials in the KS theory and also orbital free (density based) DFT. In the following we begin with a discussion of some exact results with density-functional theory.

## 1. Some exact results in DFT

In this section we discuss some important exact results of DFT, pertaining mainly to the KS framework. As mentioned above, these results provide deeper insights into DFT and are therefore useful in developing the theory further. In addition, they are helpful in understanding the results of a DFT calculation properly. Of course, the list is not exhaustive, and we have restricted our discussion to only those results that are relevant for this article. Furthermore, we will present these without going into details of their derivation.

> Exercise: A trivial exact result is the kinetic energy functional of a single electron in the ground-state. Show that this functional is $\frac{1}{8}\int \frac{|\nabla \rho(r)|^2}{\rho(r)} dr$. Furthermore, show that the non-interacting kinetic energy of two electrons in the ground-state is also given by the same expression.

### 1a. Exchange-correlation energy and potential for a single-electron system

We start our discussion by asking a question: what is the exchange-correlation energy and the corresponding potential for the electron in a hydrogen atom. One would immediately answer that it is zero because there is no interaction energy. A little more thought, however, tells us that this is not the answer. We have a density and calculate all the energy components from this density. Thus, the Hartree energy and the exchange-correlation energies are not zero separately. Yet the total energy of the electron in a hydrogen atom is the sum $\langle \varphi_i | -\frac{1}{2}\nabla^2 | \varphi_i \rangle + \int \rho(r) v_{ext}(r) dr$ of its kinetic energy and its energy of interaction with the nucleus. This immediately implies that the Hartree and exchange-correlation energies add up to zero. For a single electron system, therefore,

$$E_{xc}^{DFT}[\rho_{single}] = -\frac{1}{2} \iint \frac{\rho_{single}(r)\rho_{single}(r')}{|r-r'|} dr dr' \quad , \quad (1)$$

where $\rho_{single}(r)$ is the density of a single-electron. It follows that the exchange-correlation potential for a single electron is

$$v_{xc}^{single}(r) = -\int \frac{\rho_{single}(r')}{|r-r'|} dr' \quad . \quad (2)$$

From a physical perspective, Eqs. (1) and (2) express in the language of DFT that an electron cannot interact with itself, i.e., its **self-interaction energy** is zero. Most of the approximate exchange-correlation energy functionals, for example the LDA functional discussed in DFT-I, do not satisfy these conditions and that contributes to the error in the results obtained using these functionals.

> Question: Why can't we write the exchange-correlation energy of more than one electron as the sum of $E_{xc}^{DFT}[\rho_{single}]$ for each electron?

**1b. Fermi-Coulomb hole and its connection with Exchange-correlation energy and potential**

Now we discuss an exact result which leads to a physical understanding of the exchange-correlation energy and potential in terms of a quantity known as the **Fermi-Coulomb hole** or **exchange-correlation hole**, usually written as $\rho_{xc}(r,r')$. This hole arises [2] when expectation value of the joint probability operator $\frac{1}{N(N-1)}\sum_{i,j(i\neq j)} \delta(r-r_i)\delta(r'-r_j)$ of finding two electrons simultaneously at $r$ and $r'$ is calculated. The expectation value comes out to be $\int |\Psi(r,r';r_3,r_4\cdots r_N)|^2\, dr_3 dr_4\cdots dr_N$. We wish to write it as a product of probability of finding an electron at $r$, which is $\rho(r)/N$, and the probability of finding another electron from the rest $(N-1)$ electrons at $r'$. By doing some algebraic manipulations that involve adding and subtracting $\left(\frac{\rho(r)}{N}\right)\left(\frac{\rho(r')}{N-1}\right)$ and rearranging the resulting expression, we write the expectation value above as

$$\frac{\rho(r)}{N} \times \frac{1}{N-1}\left[\rho(r') - \left\{\rho(r') - \frac{N(N-1)}{\rho(r)}\int |\Psi(r,r';r_3,r_4\cdots r_N)|^2\, dr_3 dr_4\cdots dr_N\right\}\right]. \quad (3)$$

The term in the curly brackets in Eq. (3) above is the exchange-correlation hole $\rho_{xc}(r,r')$. As is evident, it represents the reduction in probability of finding an electron at $r'$ around another electron at $r$ in a many-electron system. Equivalently, it is the deficit in the density $\rho(r')$ around the electron at $r$, which is not a part of the rest of the $(N-1)$ particles anymore. Therefore, the total number of electrons excluded by the exchange-correlation hole is exactly one, i.e.,

$$\int \rho_{xc}(r,r')\, dr' = 1 \quad\quad . \quad (4a)$$

With this the expression in the square brackets in Eq. (3) integrates to $(N-1)$ as it should. Eq. (4a) is referred to as the **normalization condition for the exchange-correlation hole**. It is now easy to show, and is left for the reader to work out, that the exchange-correlation energy given in Eq. (68) of DFT-I is

$$E_{xc}^{QM}[\rho] = -\frac{1}{2}\iint \frac{\rho(r)\rho_{xc}(r,r')}{|r-r'|}\, dr dr' \quad\quad . \quad (5a)$$

It therefore represents the energy of interaction between the electrons in a system and the exchange-correlation charge density around each one of them.

At this stage, the reader may be wondering that all the equations above have been obtained using a wavefunction and if they would remain valid in KS formalism also. The answer is in the affirmative. Exchange-correlation energy $E_{xc}^{DFT}[\rho]$ in KS-DFT defined in Eq. (74) of DFT-I can also be written in the form of Eq. (5a) with $\rho_{xc}(r,r')$ now replaced by the exchange-correlation hole $\rho_{xc}^{DFT}(r,r')$ in DFT. This is given as Eq. (5b) in section 3c. For details on how the exchange-correlation hole is defined in DFT and how it incorporates the effects of $T_c[\rho]$, the difference between the true kinetic energy and the non-interacting kinetic energy, we refer the reader to the literature [3]. It suffices here to say that $\rho_{xc}^{DFT}(r,r')$ also satisfies the normalization condition (4a).

The exchange-correlation energy and the corresponding hole can be calculated for the exact as well as an approximate wavefunction. If the exact wavefunction is used, the exchange-correlation hole represents the effects of the Pauli-exclusion principle and electron-electron interaction. On the other hand, if we used the HF wavefunction or the Slater-determinant of KS orbitals, the exchange-correlation energy obtained is the exchange energy and the corresponding hole is the **exchange-hole**, also known as the **Fermi hole**. In the following paragraph we discuss these **exchange-only** quantities.

> Exercise: Obtain the exchange-correlation hole for a product wavefunction.

The exchange energy defined in Eq. (43) can be expressed as

$$E_x[\rho] = -\frac{1}{2} \iint \frac{\rho(r)\rho_x(r,r')}{|r-r'|} dr dr' \quad , \tag{6}$$

where the exchange hole (or the Fermi hole)

$$\rho_x(r,r') = \frac{1}{\rho(r)} \sum_{m_s} \sum_{i,j} \varphi_{i,m_s}^*(r)\varphi_{j,m_s}^*(r')\varphi_{i,m_s}(r')\varphi_{j,m_s}(r) \quad . \tag{7}$$

in terms of the orbitals. These orbitals can be either the HF orbitals or the KS orbitals. Like the exchange-correlation hole, exchange hole also gives the reduction in the probability of finding an electron at $r'$ around another electron at $r$ in a many-electron system, but only for electrons of the same z-component of spin. Furthermore, it arises due to the Pauli exclusion principle alone. Consequently, Fermi hole exists around each electron in a many-electron system irrespective of whether the electrons are interacting or non-interacting. From the expression in Eq. (7), it is easy to see that the normalization condition is satisfied by the Fermi hole also, i.e.,

$$\int \rho_x(r,r') dr' = 1 \quad . \tag{4b}$$

In addition, the exchange hole is positive throughout space, i.e.

$$\rho_x(r,r') \geq 0 \quad \text{for all } r' \quad . \tag{4c}$$

Finally, the correlation hole $\rho_c(r,r')$ is obtained by subtracting the exchange hole from the correlation hole (this follows from the definition of correlation energy). Therefore, the net charge carried by the correlation hole vanishes, i.e.,

$$\int \rho_c(r,r') dr' = 0 \quad . \tag{4d}$$

The correlation hole $\rho_c(r,r')$ has both positive and negative values as a function of $r'$ since its integral vanishes.

The next question we address is if the exchange-correlation potential can also be interpreted physically in terms of the exchange-correlation hole and as a consequence be obtained directly from it. An immediate possibility is that

$$v_{xc}(\mathbf{r}) = -\int \frac{\rho_{xc}(\mathbf{r},\mathbf{r}')}{|\mathbf{r}-\mathbf{r}'|} d\mathbf{r}' \quad . \tag{8a}$$

However, if the reader ponders over it a bit, it will become clear that this is not correct. That is because the Fermi-Coulomb hole itself is a functional of the density and in writing expression of Eq. (8) as the functional derivative of the exchange-correlation energy, this dependence has not been taken into account. The same is true for the exchange-only potential also.

> In the exchange-only case the potential of Eq. (8a) is known as the Slater potential and its expression in terms of the Fermi hole is
>
> $$v_x^{Slater}(\mathbf{r}) = -\int \frac{\rho_x(\mathbf{r},\mathbf{r}')}{|\mathbf{r}-\mathbf{r}'|} d\mathbf{r}' \quad . \tag{8b}$$
>
> We will make use of Slater potential later in the article when discussing modelling of exchange-correlation potentials.

The correct way to obtain the potential from the Fermi-Coulomb hole therefore would be [4] to calculate the work done in moving an electron in the electric field produced by the Fermi-Coulomb hole. Thus,

$$v_{xc}(\mathbf{r}) = -\int_{\infty}^{r} \boldsymbol{\mathcal{F}}(\mathbf{r}') \cdot d\mathbf{l}' \quad , \tag{9}$$

where

$$\boldsymbol{\mathcal{F}}(\mathbf{r}) = -\int \frac{\rho_{xc}(\mathbf{r},\mathbf{r}')}{|\mathbf{r}-\mathbf{r}'|^3}(\mathbf{r}-\mathbf{r}')d\mathbf{r}' \tag{10}$$

is the force on an electron due to its Fermi-Coulomb hole. Similarly, the exchange potential can be calculated using expressions above using the Fermi hole $\rho_x(\mathbf{r},\mathbf{r}')$ in place of the Fermi-Coulomb hole $\rho_{xc}(\mathbf{r},\mathbf{r}')$. In the literature, this potential is also referred to as the Harbola-Sahni (HS) potential. With the expression for potential given in Eq. (9), a question arises if it is path-independent or equivalently if the curl of the force field given in Eq. (10) zero. For symmetric systems, such as those with spherical symmetry, $\nabla \times \boldsymbol{\mathcal{F}}$ indeed vanishes. For such systems, the potential of Eq. (9) comes out to be essentially exact, as do the results [5] from self-consistent calculations employing this potential within the exchange-only theory. The potential for argon (Ar) atom thus obtained is shown in Figure (1) where it is also compared with the corresponding exact exchange potential (see section 3e below); their proximity is self-evident. Thus, the proposal above provides the physical basis of Kohn-Sham formalism by interpreting the exchange-correlation potential correctly. Further theoretical analysis [6] shows that the slight variation between the exact potential and that obtained via Eq. (9) arises from the difference in the non-interacting kinetic energy and the true kinetic energy of electrons. Adding such a term to the potential above makes it both path-independent and exact. The picture given above is derived [7] mathematically from differential virial theorem. This derivation gives the exact expression for the kinetic energy terms. This way of looking at KS theory in terms of fields has been given the name Quantal DFT [8] and provides a physical perspective to it.

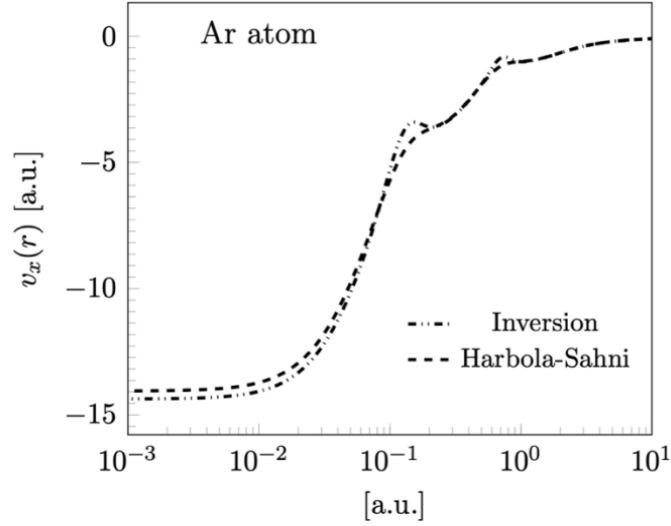

*Fig (1): The exact exchange potential of Ar atom obtained by inversion of its HF density and the potential of Eq. (9) obtained from the self-consistent solution of the corresponding Kohn-Sham equation.*

### 1c. Asymptotic behaviour of exchange-correlation energy per electron

Although the exact exchange-correlation energy functional or equivalently exchange-correlation energy per electron is not known in general, its behaviour in the regions far away from a many-electron system can be given exactly. Having this knowledge helps in designing better approximations for these functionals.

To learn about the asymptotic behaviour, we are going make use of the DFT counterpart of Eq. (5a) it is, which is

$$E_{xc}[\rho] = -\frac{1}{2}\iint \frac{\rho(r)\rho_{xc}(r,r')}{|r-r'|} dr dr' \qquad . \qquad (5b)$$

Here all the quantities are as defined in DFT. It is evident from Eq. (5b) that the exchange-correlation energy per electron is

$$\epsilon_{xc}(r) = -\frac{1}{2}\int \frac{\rho_{xc}(r,r')}{|r-r'|} dr' \qquad . \qquad (11)$$

The same expression is true for the exchange energy per electron $\epsilon_x(r)$ in terms of the Fermi hole $\rho_x(r,r')$. For distances $r$ far ($r \gg system\ size$) from a finite-sized system, such as atoms, molecules and clusters, both the exchange-correlation and exchange energy per electron are

$$\epsilon_{xc}(r) \to -\frac{1}{2r} \qquad (12a)$$

and

$$\epsilon_x(r) \to -\frac{1}{2r} \qquad (12b)$$

This is easily seen by using the normalization condition of Eqs. (4a) and (4b) satisfied by the Fermi-Coulomb and Fermi holes, respectively. When an electron is at a large distance from the system, its Fermi-Coulomb hole is left behind among the electrons in the system. Since the hole has a total charge of unity, the term $\int \frac{\rho_{xc}(r,r')}{|r-r'|} dr'$ will have the value $\frac{1}{r}$ at large distance, giving the result stated above. The same is true for $\epsilon_x(r)$.

**1d. Asymptotic behaviour of exchange-correlation potential for finite and semi-infinite systems**

Asymptotic behaviour of the exchange-correlation potential is of considerable interest since structure of the potential in outer regions of a system affects highest occupied KS orbital and those close to it significantly; and it is these orbitals that are mainly responsible for determining properties of interest of a material. Results for asymptotic behaviour of exchange-correlation potential for both finite and semi-infinite systems were first obtained [9] theoretically by employing quasi-particle amplitude of electrons in a many-electron system.

For finite systems, the exchange-correlation potential

$$v_{xc}(r) \to -\frac{1}{r} \quad (13a)$$

for distance $r$ far away from the system. Like for the exchange-correlation energy per electron, this follows from the physical origin of the potential being the field produced by the Fermi-Coulomb hole, which carries a total charge of unity. For an electron is far from the system, its Fermi-Coulomb left behind in the system hardly changes with the position of the electron. This leads to the potential energy of the electron in field being $-\frac{1}{r}$ at large distances, giving the result stated above. A classical argument for the same conclusion can also be given as follows. For an $N$ electron system, the Hartree potential far away from it has the functional form $\frac{N}{r}$. Now the electron far away can be considered to be distinct from $(N-1)$ electrons left behind and experiences a potential $\frac{N-1}{r}$, which is the sum of the Hartree potential of $N$ electrons and the exchange-correlation potential. This leads to the result of Eq. (13a). From these arguments, it is also clear that the exchange potential itself has the asymptotic dependence

$$v_x(r) \to -\frac{1}{r} \quad (13b)$$

for finite-sized systems. Thus the correlation potential decays faster than $\frac{1}{r}$ as $r \to \infty$ and its dependence on the distance is proportional to $\frac{1}{r^4}$ [9].

For semi-infinite systems like a metal filling the region $z \le 0$, the exchange-correlation potential far from the surface at $z = 0$ is [9, 10] the image potential, i.e.,

$$v_{xc}(r) \to -\frac{1}{4z} \quad . \quad (14)$$

For this result also a classical argument similar to the one given above can be made. To get the result from Fermi-Coulomb hole, however, requires a numerical calculation because the hole is highly delocalized. In addition, in case of semi-infinite systems, it is not conclusively clear [11] if exchange potential itself goes as $-\frac{1}{4z}$ in asymptotic regions although for certain values of Wigner-Seitz radii, this appears [12] to be the case.

It is easily seen that the LDA exchange-correlation or the exchange potential fails to satisfy the exact properties discussed above because for densities decaying exponentially, the potential also decays exponentially. As a result, the LDA potential is less binding than the exact potential and leads to eigen-energies for highest occupied orbitals that are substantially in error in comparison to the correct ionization potential of a system, as the values presented in DFT-I showed. Thus, the LDA does not satisfy the ionization-potential theorem of DFT. Similarly, many accurate functionals, although giving highly accurate exchange-correlation energies, do not give the corresponding potential correctly in the asymptotic regions of a system. This leads to these functionals also not satisfying the ionization-potential theorem.

**1e. Exact exchange-correlation potential for some systems**

In this sub-section, we present the exact exchange-correlation and exact exchange potentials for some small systems. These are systems for which highly accurate densities have been obtained by solving the Schrödinger equation. Obtaining the exact potentials corresponding to a given density is known as the inverse Kohn-Sham problem and is solved by using various methods [13] developed for it. For recent reviews of these methods, see Refs. [14,15]. A numerical package *n2v* [16] that does such an inversion is also available [17]. Using these methods, exact exchange-correlation potentials can be obtained for any system. We present some of these results in the following.

In Fig. (1), we have plotted the exact exchange potential of the argon (Ar) atom obtained by inverting its Hartree-Fock density [18]. It is evident from the figure that the exact exchange potential is negative throughout the space and becomes zero in the low-density region far away from the nucleus. It is constant in the deep interior region near the nucleus and has bumps in the inter shell regions. We also note that in the asymptotic region, the exchange potential follows the exact asymptotic behaviour and goes as $-\frac{1}{r}$ with the distance $r$. From the figure we also learn that the potential calculated using Eq. (9) differs from the exact potential in the inter-shell regions where where it is flat in comparison to the exact potential.

In Fig. (2), we display the correlation potential for two two-electron spherical systems, viz., the hydrogen (H) anion and the Hookium atom. These are calculated by first inverting the density to obtain the exchange-correlation potential. Since the exact exchange potential for these systems is negative of the half of their Hartree potential, the correlation potential is obtained by subtracting it from the corresponding exact exchange-correlation potential. For the H anion, we used highly accurate density [19] calculated from 204-parameter correlated-wavefunction, and for the Hookium atom, the exact density [20] corresponding to harmonic oscillator potential $\frac{1}{2}kr^2$ with $k = 0.25$ is used.

It is evident from the figure that the correlation potential for these systems has both positive and negative values. For H anion correlation potential is negative near the nucleus, it keeps increasing up to a distance of 2 a.u. from the nucleus and then starts decreasing, becoming zero far away from the nucleus. In contrast to the H anion, the behavior of correlation potential is opposite for the Hookium atom. It is positive near the center of the atom and has well-like structure little away from the center of the atom. This structure of the correlation potential, oscillating between positive and negative values, is quite general and is seen [21] in other atoms too.

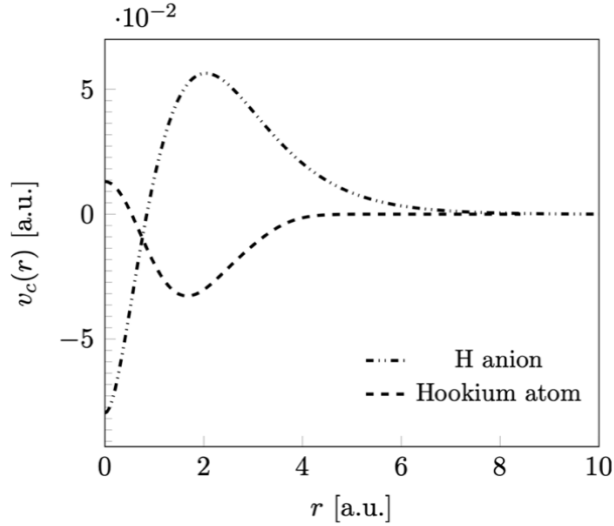

*Fig 2: Correlation potential of correlated two-electron systems- H anion and Hookium atom.*

For the systems considered above, we also demonstrate the ionization potential theorem [22] (see Eq. 82 of DFT-I) by comparing the negative of the eigenvalue $-\epsilon_{max}$ of the highest occupied orbitals corresponding to the exact Kohn-Sham potential obtained by inverting the true density for the H anion and the Hookium with the corresponding true ionization potential. For the Ar atom, the eigenvalue is compared with the corresponding value in Hartree-Fock theory since the density inverted is the Hartree-Fock density. Comparison is also made of the eigenvalue obtained by doing a self-consistent calculation using the Harbola-Sahni exchange-only potential. These quantities are displayed in Table I. It is evident that for the systems considered above, the highest occupied Kohn-Sham eigenvalue is very close to the true ionization potential and the ionization potential theorem is satisfied with high accuracy.

The satisfaction of the ionization potential theorem has been shown here for three spherical symmetric systems. In the next section we prove the theorem in general by discussing density-functional theory for fractional number of electrons [22]. Density functional theory for fractional number of electrons is also important to understand the band-gap calculation in DFT .

| SYSTEM | $-\epsilon_{max}$ | EXACT ION. POT. |
|---|---|---|
| H ANION | 0.0277 | 0.0267 |
| HOOKIUM ATOM | 1.2531 | 1.25 |
| AR (HF DENSITY) | 0.5899 | 0.5910 |
| AR (HARBOLA-SAHNI) | 0.5891 | 0.5910 |

*Table I: The highest occupied Kohn-Sham orbital eigenvalue and chemical potential for the exact density of different systems. For the Ar atom, the chemical potential is taken to be its HF value.*

Exact exchange-correlation potential has also been obtained for one-dimensional systems [23, 24], molecules [25, 26] and solids using their Hartree-Fock [27] or nearly exact densities [28] and that has led to insights into the nature of exact KS systems for these systems.

**1f.  DFT for non-integer number of electrons and ionization potential theorem**

In developing DFT, one minimizes the energy of a system by taking the functional derivative of the energy functional and equating it to a constant.  That constant is then shown to be the chemical potential (see Eq. (70b) and related discussion in DFT-I).  Since the chemical potential is the derivative of energy with respect to the number of electrons, in showing its equality to the functional derivative of the energy it is implicitly assumed that ground-state energy can be defined for non-integer number of electrons.  This assumption, however, must be put on proper foundations.  In this subsection we discuss how DFT is extended [22] to fractional number of electrons. The result is quite simple and elegant.  In addition, it has important implications for DFT description of properties of materials such as band-gaps of semiconductors and charge transfer reactions.  This theory is presented in detail below.

In quantum mechanics, a pure state has integer number of electrons.  A system can, however, contain non-integer numbers if its eigenstates are mixed statistically to construct an appropriate density-matrix. In the context of DFT, we therefore get ground-state energy for $Z$ number of electrons, where $Z$ could be an integer or a non-integer, by mixing ground-states energies of different number of electrons with a probability.  Let the ground-state energy of $I$ electrons be $E_I$ and let it have the probability $p_I$ $(0 \leq p_I \leq 1)$ in the statistical mixture of different number of electrons.  Then the following two conditions are satisfied by these quantities

$$\sum_I p_I = 1 \quad , \qquad (15a)$$

$$\sum_I p_I I = Z \quad . \qquad (15b)$$

The corresponding ground-state energy is

$$E_Z = \sum_I p_I E_I \quad , \qquad (16)$$

and its variation with respect to the arbitrary variations $\delta p_I$ in probabilities is

$$\delta E_Z = \sum_I \delta p_I E_I \quad . \qquad (17)$$

This variation should be zero for the ground-state energy. However, all variations $\{\delta p_I\}$ in Eq. (17) are not arbitrary because the two conditions above imply that probabilities for two different number of electrons are given in terms of other probabilities. We therefore first write $\delta E_Z$ in terms of all $\{\delta p_I\}$ except for the number of electrons $L$ and $M$ $(M > L)$. Then using Eqs. (15) and (16), we get

$$p_L = \frac{M-Z}{M-L} + \sum_{I(I \neq L,M)} \frac{I-M}{M-L} p_I \quad , \tag{18a}$$

and

$$p_M = \frac{Z-L}{M-L} - \sum_{I(I \neq L,M)} \frac{I-L}{M-L} p_I \quad . \tag{18b}$$

This gives the energy to be

$$E_Z = \left( \frac{M-Z}{M-L} + \sum_{I(I \neq L,M)} \frac{I-M}{M-L} p_I \right) E_L + \left( \frac{Z-L}{M-L} - \sum_{I(I \neq L,M)} \frac{I-L}{M-L} p_I \right) E_M + \sum_{I(I \neq L,M)} p_I E_I \quad , \tag{19}$$

The corresponding variation in the energy is

$$\delta E_Z = \sum_{I(I \neq L,M)} \left( \frac{I-M}{M-L} E_L - \frac{I-L}{M-L} E_M + E_I \right) \delta p_I \quad , \tag{20}$$

Now for $\delta E_Z$ to vanish for arbitrary variations in $\delta p_I$, the expression in the brackets must be zero for each $I$. However, energies $E_L$, $E_M$ and $E_I$ for number of electrons $L, M$ and $I$ are not related in a manner that will make this expression zero. Therefore, $\delta p_I$ $(I \neq L, M)$ must all be zero in order that $\delta E = 0$. Since $\delta p_I$ $(I \neq L, M)$ can be taken to be arbitrary in both magnitude and sign if $0 < p_I < 1$, the condition that all $\delta p_I$ $(I \neq L, M)$ vanish demands $p_I$ $(I \neq L, M)$ should **all** be either 1 or 0; this is the only way it can happen. To understand it in a perspicuous manner, let us ask the following question. What should be the values of the variable $p_I$ in the range $0 \leq p_I \leq 1$ about which it cannot be bot decreased or increased arbitrarily? The answer is clearly $p_I = 0$ or $p_I = 1$. The possibility of $p_I$ being 1 is obviously ruled out because of condition (15a). Hence all $p_I$ $(I \neq L, M) = 0$ and the only probabilities that are nonzero are $p_L$ and $p_M$.

> The same conclusion can also be reached with a slightly different argument. The condition that $\delta p_I$ $(I \neq L, M)$ are all zero also gives $\delta p_L = 0$ and $\delta p_M = 0$ by Eqs. (18a) and (18b). This in turn implies that one cannot vary any of the $p_I$ and they are fixed for a given number of electrons. Since all $p_I$ satisfy two conditions given by Eqs. (15a) and (15b), a unique solution emerges in the range [0,1] only if all except two of these are taken to be zero. Choice of only one $p_I$ being non-zero makes the problem unsolvable and taking more than two of them being nonvanishing gives many possibilities for different $p_I$ so that we can have $\delta p_I \neq 0$ and still satisfy Eqs. (15a) and (15b).

Therefore

$$p_L + p_M = 1 \quad , \tag{21a}$$

$$p_L L + p_M M = Z \quad , \tag{21b}$$

and

$$E = p_L E_L + p_M E_M \quad . \tag{22}$$

Next question that arises is: how do we fix $L$ and $M$? To start with, $L$ and $M$ must be consecutive integers. If they are not and $M = L + n$, where $n$ is an integer, the energy of number of electrons $X = L + m$ ($m < n$), where $m$ is also an integer, will turn out to be

$$E_{L+m} = \left(1 - \frac{m}{n}\right) E_L + \frac{m}{n} E_{L+n} \quad . \tag{23}$$

Question: Why should the number of electrons $X$ lie between $L$ and $M$?

However, energies for number of electrons $L$, $L + m$ and $L + n$ are not necessarily related as given in Eq. (23). For example, for an $N$-electron system, take $L = N - 1, M = N + 1$ so that $n = 2$. Then it given the energy of the $N$-electrons ($m = 1$) to be $E_N = (E_{N-1} + E_{N+1})/2$, which cannot be true in general. This spurious result does not arise if $n = 1$ and $M = L + 1$. Consequently, probabilities for number of electrons $Z = L + f$ ($0 \leq f \leq 1$) are $p_L = (1 - f)$ and $p_{L+1} = f$, which give

$$E_{L+f} = (1 - f) E_L + f E_{L+1} \quad . \tag{24}$$

Question: What is the electronic density $\rho_{L+f}(r)$ for number of electrons $Z = L + f$ ($0 \leq f \leq 1$) in terms of the densities $\rho_L(r)$ and $\rho_{L+1}(r)$ for $L$ and $L + 1$ electrons, respectively?

Result of Eq. (24) is very simple. It gives the energy of fractional number of electrons by connecting the energies of adjacent integers by straight line segments. This is shown in Figure 3 for $N - 1, N$ and $N + 1$ electrons in a system. The slope $\frac{dE}{dZ}$ and therefore the chemical potential of the system is

$$\mu = \begin{cases} -I & \text{for} \quad (N - 1) < Z < N \\ -A & \text{for} \quad N < X < (N + 1) \end{cases}, \tag{25a}$$

where $I = E_{N-1} - E_N$ and $A = E_N - E_{N+1}$ are, respectively, the ionization potential and electron affinity for the $N$-electron system.

A question that arises naturally here is what is value of $\mu$ as $Z$ becomes equal to $N$. Let us first consider the case of $Z$ approaching $N$ from below. In this case $\mu = -I_N$, where $I_N$ is the ionization potential $I$ of the $N$-electron system. The reason is that $\mu$ for $(N - 1) < Z < N$ is related to the ionization potential of the $N$-electron system so will remain unchanged and equal to $-I_N$ even as $Z$ becomes equal to $N$ reaching it from below; and as soon as $Z$ crosses $N$, $\mu$ will change discontinuously to $-I_{N+1}$, i.e., negative of the ionization potential of $(N + 1)$-electron system or equivalently, $-A$ which is the negative of the electron affinity of $N$-electron system. On the other hand, if $Z$ approaches $(N - 1)$ from above, $\mu = -I_N$ and will change to $-I_{N-1}$ as soon as $Z$ crosses $(N - 1)$ and varies in the range $(N - 2) < Z < (N - 1)$. Combining the two arguments above, Eq. (25) can be rewritten as

$$\mu = \begin{cases} -I & \text{for} \quad (N - 1) < Z \leq N \\ -A & \text{for} \quad N < Z \leq (N + 1) \end{cases}. \tag{25b}$$

> Notice the equality sign on the right limit of $Z$. Thus, the chemical potential is an electronic system is in general equals negative of its ionization potential (electron affinity) if an electron is being removed from (added to) it. The average chemical potential
>
> $$\mu_{average} = -\frac{I+A}{2} \quad , \tag{26}$$
>
> of a system and its relative value in comparison with $\mu_{average}$ of another system will therefore govern [23] the exchange of electrons between the two. It is not surprising then that $\mu_{average}$ of an electronic system is equal [29] to negative of its Mulliken electronegativity [30]. Thus, DFT helps define an important chemical concept precisely. Further studies along these lines have given rise to conceptual DFT [31] that has helped clarify many important chemical concepts.

We summarize that the equality of the chemical potential and ionization potential of a system has been proved here rigorously by extending DFT to include electron densities corresponding to fractional number of electrons.

Before we use the results above to prove the ionization potential theorem, we investigate if the energy $E_{L+f}$ obtained in Eq. (24) is a minimum for the number of electrons $L+f$. For this, let us consider an $N$-electron system with fractional number of electrons $N+f$ so that the probability of having different number of electrons in the system is $p_{N-1}=0$, $p_N=1-f$ and $p_{N+1}=f$. As elaborated in discussion after Eq. (20), the energy change $\delta E \neq 0$ when $\delta p_I$ $(I \neq N, N-1) \neq 0$. We now show that this change is positive so that the energy given by Eq. (24) is indeed the minimum possible for number of electrons $N+f$. Here we take that change to be $\delta p_{N-1} > 0$ in $p_{N-1}=0$ and use Eq. (20) with $I = N-1$, $L = N$ and $M = N+1$ to get the corresponding change in the energy

$$\delta E = (-2E_N + E_{N+1} + E_{N-1})\delta p_{N-1} \quad . \tag{27}$$

Since the $E(N)$ versus $N$ curve is convex, expression in the parenthesis above is positive. This gives $\delta E > 0$ because $\delta p_{N-1} > 0$ and shows that energy for fractional number of electrons given by Eq. (106) is indeed a minimum with respect to variations in $p_I$. Convexity of the $E(N)$ versus $N$ curve means that the ionization potential $I = E_{N-1} - E_N$ of a system is greater than its electron affinity $A = E_N - E_{N+1}$, which is indeed observed in nature.

> Repeat the exercise above for the number of electrons $N-f$ taking $\delta p_{N+1} > 0$. Also show that if $\delta p_I$ is taken to be non-zero for some other values of $I$, convexity of $E(N)$ versus $N$ graph always gives $\delta E > 0$.

To summarize, if $E(N)$ satisfies convexity condition $E_{N+1} - 2E_N + E_{N-1} > 0$, minimum energy for fractional number of electrons is obtained uniquely by mixing appropriately the energies for the integer number of electrons adjacent to that fraction, as given by Eq. (24). In essence the result indicates that to be consistent with energies for integer number of electrons and their convexity, there is only one choice for mixing states with different integer number of electrons to obtain the minimum energy for a given non-integer number of electrons.

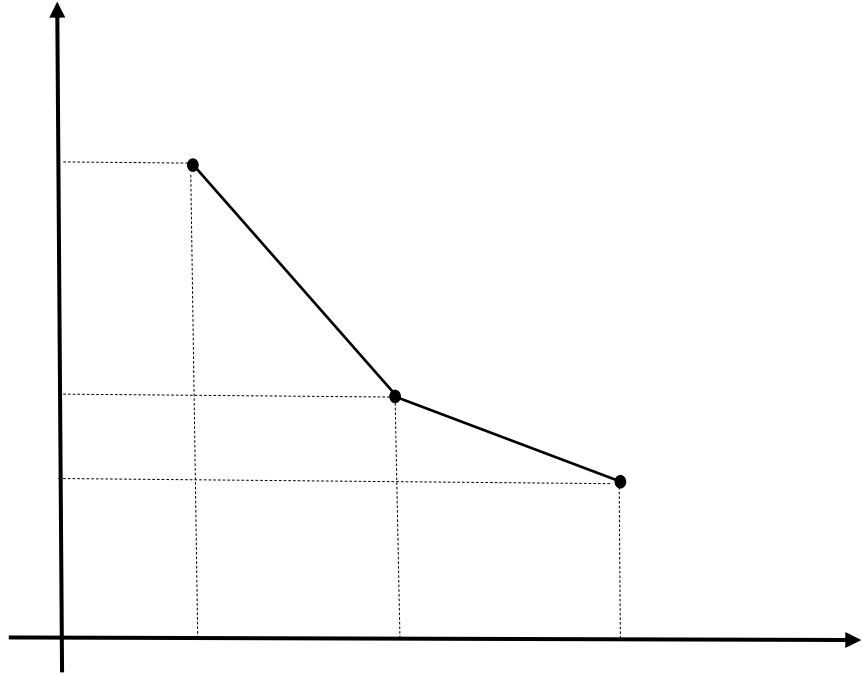

***Fig. (3):*** *Energy $E_X$ plotted against the number of electrons $X$ ($N-1 < X < N+1$). Notice the linear segments that describe the change in energy with electron number between two integers.*

We now prove the ionization potential theorem, stated earlier in Eq. (82) of DFT-I. For electron number $Z = N - 1 + f$ ($0 < f \leq 1$), the energy of the KS system calculated using Eq. (81) of DFT-I and Eq. (24) above is

$$E_{N-1+f} = \sum_1^{N-1} \epsilon_i + f\epsilon_N \quad . \tag{27}$$

This gives for $(N-1) < Z \leq N$ (discussion in the box after Eq. 25 to explains the equality sign on the right limit of $Z$ for the Kohn-sham system too)

$$\mu = \epsilon_N$$
$$= \epsilon_{max} \quad \text{for } N \text{ electron system} \quad . \tag{28}$$

As was noted earlier, chemical potential of the true system and that of the KS system are the same because of the way the KS system is constructed. This combined with Eqs. (25b) and (28) leads to the **ionization potential (IP) theorem** in a rigorous manner. These results also imply that for KS calculations with the exact energy functionals, the eigenenergy of the highest occupied orbital will remain constant for the number of electrons varying between two integers and including the larger integer.

Besides being an exact result of fundamental significance, Ionization potential theorem has also been used in fixing [32] theoretically a free parameter in some exchange-correlation energy functionals. This is done as follows. One performs a KS calculation with the functional in use for $N$ and $N-1$ electrons for different values of the parameter in the functional. The value of this parameter is then fixed by satisfying $E_{N-1} - E_N = -\epsilon_N$, where $\epsilon_N$ is the highest occupied orbital eigenenergy for the $N$ electron system.

> Calculate energy $E(N)$ for fractional number of electrons $N$ using the LDA/LSD exchange-correlation energy functional. The density for fractional number of electrons can be obtained either by a KS calculation with fractional occupancy of orbitals or by mixing densities for integer number of electrons calculated from known wavefunctions. Does the resulting curve look like that in Figure 3?

**1g. Derivative discontinuity of the exchange-correlation potential**

A subtle consequence of the discussion in subsection (3e) is that the exchange-correlation potential of a system changes by a constant when the number of electrons in it passes through an integer. This is a consequence of the chemical potential changing by a constant across an integer number of electrons, as is evident from Eq. (25) and Figure 3. We first show how such a discontinuity leads to a change in the functional derivative of the kinetic energy for a non-interacting system. We then move on to interacting systems and discuss that such discontinuous changes also occur in the exchange-correlation potential of an interacting system when electron number passes through an integer. Furthermore, this discontinuity explains why fundamental gaps of insulators and semiconductors are underestimated in a KS calculation.

Consider a system of $N$ non-interacting electrons moving in the potential $v_{KS}(\boldsymbol{r})$ with its highest occupied orbital fully filled. The chemical potential of the system is $\epsilon_N$ for number of electrons between $N-1$ and $N$ and it is $\epsilon_{N+1}$ for number of electrons between $N$ and $N+1$; we point out that the orbital eigenvalues in a system of non-interacting electrons remain the same irrespective of the number of electrons in it. Thus, if there is no degeneracy in the $N$-electron system, the chemical potential changes by a constant $\epsilon_{N+1} - \epsilon_N$ as the electron number crosses $N$. Since the potential for non-interacting electrons is fixed, Eq. (72) then implies:

$$\left.\frac{\delta T_s[\rho]}{\delta \rho(\boldsymbol{r})}\right|_{N+f} - \left.\frac{\delta T_s[\rho]}{\delta \rho(\boldsymbol{r})}\right|_N = \epsilon_{N+1} - \epsilon_N \quad . \quad (29)$$

where $f$ is infinitesimally small. Eq. (29) clearly shows that the functional derivative of the non-interacting kinetic energy will change by a constant greater than or equal to zero as the electron number passes through an integer. The change will be zero in the case of the highest orbital being degenerate and not being fully occupied.

> Consider non-interacting electrons in the potential $-\frac{Z}{r}$. If the number of electrons changes from 2 to 3, what will be the corresponding change in the functional derivative of their kinetic energy? What will be this change for electron number increasing from 3 to 4?

Let us now look at what happens when electrons are interacting. In this case, as a fraction of an electron is added to the system, the corresponding exchange-correlation potential changes discontinuously by a constant throughout the system except in regions asymptotically far from it, where the potential goes to zero. This is best illustrated by the example of Helium atom and its positive ion. So, we first discuss this example and then present the general result. In both these systems, electrons occupy the degenerate $1s$ orbital. The ionization potential of Helium cation is 2.0 atomic units (au). while its electron affinity (ionization potential of Helium atom) is 0.9 au. So, the chemical potential of Helium cation jumps by 1.1 au as the electron number crosses 1. By the IP theorem, therefore, the eigenvalue $\epsilon_{1s}$ must change from $-2.0$ au to $-0.9$ au as the electron number changes from 1 to $1 + f$ with $f \to 0$. On the basis of this, we conclude from Eqs. (75), and (78) of DFT-I that the exchange-correlation potential would also change by 1.1 au in this process. This is elaborated upon in the following paragraphs.

Looking at Eq. (75) of DFT-I, when electron number crosses 1, the functional derivative of the kinetic energy does not change because of the degeneracy of 1s orbital and the Hartree potential hardly changes because the change in the density of electrons is insignificant. Therefore, the discontinuous constant change in the chemical potential can arise only if the exchange-correlation potential changes by the same amount. Let us now see how we get the same result from the KS equation.

For the Helium atom with $1 + f$ electrons, the density of electrons is

$$\rho_{1+f}(r) = (1 - f)\rho_1(r) + f\rho_2(r) \quad , \tag{30}$$

where $\rho_1(r)$ and $\rho_2(r)$ are the electron densities of He cation and He atom, respectively. Corresponding to this density, there is only one KS orbital

$$\varphi_{1+f}(r) = \sqrt{\frac{(1 - f)\rho_1(r) + f\rho_2(r)}{1 + f}} \quad , \tag{31}$$

with the orbital energy $-0.9$ au by the IP theorem. For $f \to 0$, the orbital above is that corresponding to $\rho_1(r)$, i.e., it is essentially the same as the orbital for the electron the He ion. Therefore, when substituted in the KS equation, its eigenvalue (in atomic units) should have been $-2.0$ instead of $-0.9$. Since the Hartree potential for $\rho_{1+f}(r)$ is hardly different from that for $\rho_1(r)$, the only way the orbital energy for $\varphi_{1+f}(r)$ can be $-0.9$ au is if the exchange-correlation potential changes by a constant equal to 1.1 a.u. Thus, the exchange-correlation potential can change by a constant as the electron number passes through an integer.

Looking at the conclusion above, one may ask: if the potential jumps by a constant, how can it go to zero as $r \to \infty$? To answer this question, let us calculate the exchange-correlation potential $v_{xc}^{1+f}(\vec{r})$ for the He ion with $1 + f$ electrons by inverting the KS equation. It is given as

$$v_{xc}^{1+f}(r) = \frac{1}{2}\frac{\nabla^2 \varphi_{1+f}(r)}{\varphi_{1+f}(r)} - \frac{2}{r} - \int \frac{\rho_{1+f}(r')}{|r - r'|} dr' + \text{constant} \quad , \tag{32}$$

where the constant is fixed by the boundary condition on the potential; we fix it as

$$\lim_{r \to \infty} v_{xc}^{1+f}(r) \to 0 \quad . \tag{33}$$

To understand the structure of $v_{xc}^{1+f}(r)$, we now look at the behaviour of the density in the asymptotic regions of a system. In the limit of $r \to \infty$ the ground-state density of a system has [33] the functional dependence $e^{-2\sqrt{2I}r}$. As a result, in the case of He considered above

$$\lim_{r \to \infty} \begin{cases} \rho_1(r) \sim e^{-4r} \\ \rho_2(r) \sim e^{-2\sqrt{1.8}r} \end{cases} . \qquad (34)$$

As is clear from Eq. (34), while density $\rho_{1+f}(r)$ is essentially equal to $\rho_1(r)$ within the system and it becomes $f\rho_2(r)$ in the asymptotic regions. This gives $\varphi_{1+f}(r) \cong \varphi_1(r)$ inside the atom. Here $\phi_1(r) = \sqrt{\rho_1(r)}$ is the KS orbital for the electron in the He ion. On the other hand, $\varphi_{1+f}(r) \cong \sqrt{f}\phi_2(r) \sim e^{-\sqrt{1.8}r}$ as $r \to \infty$, with $\phi_2(r) = \sqrt{\rho_2(r)/2}$ being the KS orbital for the He atom that has two electrons in it. Using these together with the boundary condition of Eq. (33) gives constant in Eq. (32) to be $-2.0$ au for $v_{xc}^1(r)$. Thus

$$v_{xc}^1(r) = \frac{1}{2}\frac{\nabla^2 \varphi_1(r)}{\varphi_1(r)} - \frac{2}{r} - \int \frac{\rho_1(r')}{|r-r'|}dr' - 2.0 \qquad , \qquad (32a)$$

inside $He^+$. On the other hand, the constant will be $-0.9$ au for $v_{xc}^{1+f}(r)$. In the limit of $f \to 0$, this gives

$$v_{xc}^{1+f}(r) = \frac{1}{2}\frac{\nabla^2 \varphi_1(r)}{\varphi_1(r)} - \frac{2}{r} - \int \frac{\rho_{1+f}(r')}{|r-r'|}dr' - 0.9 \qquad . \qquad (32b)$$

inside the ion with (1+f) electrons. Eqs. (32a) and (32b) lead to

$$v_{xc}^{1+f}(r) - v_{xc}^1(r) = 1.1 \qquad (32c)$$

for $f \to 0$, since the difference in the Hartree potential is insignificant in this limit. Thus, inside the ion where $\varphi_{1+f}(r) \cong \varphi_1(r)$ the exchange-correlation potential changes by a constant (equal to the difference between the ionization potential and electron affinity of the He ion in this case). On the other hand, both the potentials go to zero as $r \to \infty$, so that the difference is zero in the limit. The transition of the difference $v_{xc}^{1+f}(r) - v_{xc}^1(r)$ from a being a constant to it becoming zero takes place in the region where the orbital $\varphi_{1+f}(r)$ changes from to $\varphi_1(r)$ to $f\phi_2(r)$. Thus, smaller the value of $f$, farther out from the nucleus will be the region where the transition takes place. The results presented above have been shown [34-36] explicitly by numerical calculations.

> Using Eqs. (31), (32a) and (32b), obtain the exchange-correlation potentials for $\rho_1(r)$ and $\rho_{1+f}(r)$ and the difference between them for several values of $f$ to confirm the conclusions arrived at above. Density for the He atom can be taken from Koga et al. [37]

In the above, we have demonstrated through two-electron system that the exchange-correlation potential jumps by a constant. This is known as the derivative-discontinuity of the exchange-correlation potential and will be denoted as $\Delta_{xc}$ in the following. The question that arises now is if $\Delta_{xc}$ will be nonzero in general for all many-electron systems. The answer is in the affirmative. This is so since there is no reason for the change in the chemical potential to be equal only to the difference in the eigenvalues for the lowest unoccupied and highest occupied orbitals, which are eigen-energies for a virtual non-interacting system. The next question is if $\Delta_{xc}$ will always be positive. Let us see why this should be the case. Since (i) the chemical potential for the system with $(N+f)$ electrons is equal to that of $(N+1)$ electrons, as given in Eq. (25b) above, and (ii) the Coulomb interaction between electrons makes the highest occupied eigenvalue $\epsilon_{N+1}(N+1)$ for ($N + 1$) electron system more positive that the lowest unoccupied eigenvalue $\epsilon_{N+1}(N)$ for the corresponding $N$ electron system, one would have $[\epsilon_{N+1}(N + 1) - \epsilon_N(N)] > [\epsilon_{N+1}(N) - \epsilon_N(N)]$. Of course, the difference between $\epsilon_{N+1}(N + 1)$ and $\epsilon_{N+1}(N)$ will include the effect of Coulomb interaction on all components of the energy. Another way to see that it will be greater than zero is through differential virial theorem for fractional number of electrons [34,35]. In Ref. [38], it is further shown that the derivative discontinuity is solely due to the change in the kinetic energy due to electron-electron interaction.

To conclude the discussion above, it has been shown that the exchange-correlation potential changes discontinuously by a constant as the electron number increases from an integer $N$ to ($N + f$), where $f \ll 1$, in regions $(1 - f)\rho_N(\mathbf{r}) \gg f\rho_{N+1}(\mathbf{r})$. In the asymptotic regions the difference vanishes. The transition from being a constant to slowly going to zero takes place in the regions where $(1 - f)\rho_N(\mathbf{r}) \sim f\rho_{N+1}(\mathbf{r})$. Therefore, smaller the value of $f$, farther out in the asymptotic region will this transition take place.

**1h. Derivative discontinuity of $v_{xc}$ and the band-gap of semiconductors**

An important consequence of the derivative discontinuity of the exchange-correlation potential is that the KS gap will always be smaller [39,40] than the true fundamental gap - defined as $(I - A)$ with $I$ and $A$ being, respectively, the ionization potential and electron affinity - of a semiconductor or an insulator not only in the LDA (see Table 5 of DFT -I) but also in the exact KS theory. This is shown schematically in figure 4 where on the left we have shown the KS eigenvalues $\epsilon_N(N)$ and $\epsilon_{N+1}(N)$, respectively, for the top of the valence band and the bottom of the conduction band for a semiconductor with valence band filled with $N(N \to \infty)$ electrons and conduction band being empty. The KS gap for the system is $E_g^{KS} = \epsilon_{N+1}(N) - \epsilon_N(N)$. As an electron is added to the system taking the number of electrons to $N + 1$ (since $N \to \infty$, adding one extra electron to the system is equivalent to adding a fractional electron in a system with small value of $N$), the exchange-correlation potential jumps by $\Delta_{xc}$. The electron is now at the bottom of the conduction band which now has the eigenvalue $\epsilon_{N+1}(N + 1) = \epsilon_{N+1}(N) + \Delta_{xc}$ ; the top of the valence band also shifts by the same amount and has the eigenvalue $\epsilon_N(N + 1) = \epsilon_N(N) + \Delta_{xc}$ . The fundamental gap in terms of the total energy is $(E_{N-1} - E_N) - (E_N - E_{N+1})$. By the IP theorem this is equal to

$$\epsilon_{N+1}(N + 1) - \epsilon_N(N) = \epsilon_{N+1}(N) - \epsilon_N(N) + \Delta_{xc}$$
$$= E_g^{KS} + \Delta_{xc} \quad . \tag{33}$$

It is evident that the fundamental gap will always be larger than the KS gap by $\Delta_{xc}$, even if an exact KS calculation could be performed. This then explains why the LDA band-gap is also [41] smaller than the true band-gap.

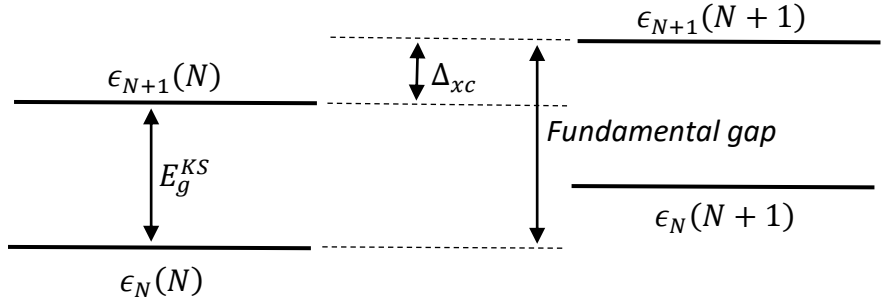

***Fig. (4):*** *On the left are shown the Kohn-Sham energy levels with eigenvalues $\epsilon_N(N)$ and $\epsilon_{N+1}(N)$ corresponding, respectively, to the top of the valence band and bottom of the conduction band for a system with its valence band filled completely with N electrons and conduction band empty. When one more electron is added to the system, the energy levels shift up by $\Delta_{xc}$ and the eigenvalues become $\epsilon_N(N+1)$ and $\epsilon_{N+1}(N+1)$, as shown on the right. The KS gap for the system is $E_g^{KS} = \epsilon_{N+1}(N) - \epsilon_N(N)$ while the fundamental gap is $\epsilon_{N+1}(N+1) - \epsilon_N(N)$.*

Finally, we comment on the difference between the optical gap and fundamental gap in a semiconductor. This is important since in many investigations, the KS gap calculated with accurate potentials comes out to be close to the optical gap. The optical gap is the difference between an excited-state energy and the ground-state energy while the fundamental gap is the difference between two ground-state properties. The former refers to the energy required to excite an electron from the top of the valence bad to the bottom of the conduction band. Thus, it can be expected on the basis of a similar process in atomic calculations [42] that calculations with accurate KS potentials may give KS gaps close to the optical gap of a semiconductor. On the other hand, the ionization potential $I$ and electron affinity $A$ of a semiconductor are measured by angle resolved photoelectron [43,44] and angle resolved inverse photoemission [45] spectroscopy and their difference gives the fundamental gap. As discussed above, this gap will always be underestimated even if the exact KS potential is employed in the calculations

## 2. Making exchange and correlation energy functionals accurate

In DFT-I, we have discussed results of DFT calculations obtained by using the LDA functional. We now turn our attention to going beyond the LDA and getting more accurate exchange and correlation energy functionals. This has been an active area of research since the beginning of DFT and continues to be so. As a result, the list of functionals developed is long and all functionals cannot be discussed here. In this section we discuss briefly some of these functionals and how they have been developed using some exact results of DFT presented above. The motivation is to make the reader familiar with methods employed to develop new accurate functionals.

### 2a. The gradient expansion approximation (GEA)

The first attempt [46] in the direction of making the energy functional more accurate was to include terms dependent on the gradient of the density in the exchange-correlation energy. The functional so developed is known as the gradient-expansion-approximation (GEA) and has the following form:

$$E_{xc}^{GEA}[\rho] = E_{xc}^{LDA}[\rho] + C_{xc} \int \frac{|\nabla\rho|^2}{\rho^{4/3}} d\boldsymbol{r} \quad , \tag{34}$$

where $C_{xc}$ is a constant. Form given above was first obtained [46] for the correlation energy and later proposed [47] on dimensional grounds for the exchange energy. The coefficient $C_c$ for the correlation energy was calculated theoretically while $C_x$ for the exchange energy was initially determined [47] by minimizing the energy of the system of interest. Later the GEA for exchange was derived [48] from first principles and value of $C_x$ was also obtained on theoretical grounds. Further refinements [49,50] and better understanding of the derivation led to slightly larger value of $C_x$ than that obtained earlier. Although incorporating the gradient correction in the energy functional makes it more accurate, the functional itself lacks in many properties of the exact functional. In the following we present this in the context of exchange energy functional. The GEA for correlation energy also follows a similar path, although it requires more complicated calculations, and we refer the interested reader to the relevant literature.

First it is easily seen that asymptotically far from a finite system, the exchange energy per electron still decays exponentially rather than as its expected exact behaviour given by Eq. (12b). Secondly, the exchange potential $v_x^{GEA}(r) = \frac{\delta E_x^{GEA}[\rho]}{\delta \rho(r)}$ becomes exponentially large in the asymptotic regions where $\rho(\boldsymbol{r}) \sim e^{-\alpha r}$ with $\alpha$ being a constant that depends on the ionization potential of the system. Finally, the exchange hole corresponding to the GEA functional does not obey [51] Eq. (4c), i.e., it is not positive at all places. In addition, its integral is not definite [51] as its value oscillates around zero without damping as a function of $r'$. Enforcing these exact properties on $E_x^{GEA}[\rho]$ improves it and leads to a new class of functionals known as generalized gradient approximation (GGA) functionals discussed next.

> Calculate the exchange-correlation potential for the GEA functional of Eq. (34) and show that it grows exponentially in asymptotic regions of a system where $\rho(\boldsymbol{r}) \sim e^{-\alpha r}$

### 2b. Generalized gradient approximation (GGA)

In GGA density and its gradient are used to write the energy functionals satisfying certain exact properties. Using the dimensionless parameter $s = \left[\frac{|\nabla\rho(r)|}{2(3\pi^2)^{1/3}(\rho(r))^{4/3}}\right]$, known as the reduced gradient, the general form of the GGA functional is usually written as (here we show it for the exchange energy)

$$E_X^{GGA}[\rho] = \int F_X^{GGA}(s)[\rho(\boldsymbol{r})\epsilon_X^{unif}(\rho(\boldsymbol{r}))] d\boldsymbol{r} \quad . \tag{35}$$

Here $F_X^{GGA}(s)$ enhances the exchange energy for a given $\rho(\boldsymbol{r})$ over the LDA. For $\nabla\rho(r) = 0$, the enhancement factor $F_X^{GGA}(s) = 1$, which reproduces the LDA exchange energy. The choice of $F_X^{GGA}$ distinguish different functionals. We describe two of these below.

***Becke exchange energy functional***: As mentioned above, the GEA functional does not give the correct behaviour of the exchange energy $\epsilon_x(\boldsymbol{r})$ per electron in the asymptotic regions of a system. To correct this, Becke incorporated a one parameter term in the GEA expression so that for exponentially decaying densities, the exact form given by Eq. (12b) is obtained. The Becke exchange energy functional [52] is

$$E_x^{Becke}[\rho] = E_x^{LDA}[\rho] - \beta \sum_{m_s} \int \rho_{m_s}^{4/3}(r) \frac{x_\sigma^2}{1+6\beta x_{m_s}\sinh^{-1}(x_{m_s})} d\boldsymbol{r} \quad , \tag{36}$$

where $\beta$ is a constant to be determined, $\rho_\sigma(\boldsymbol{r})$ is the electron density for electrons with z-component of spin $m_s$ and $x_{m_s} = \frac{|\nabla \rho_{m_s}|}{\rho_{m_s}^{4/3}}$. Value of parameter $\beta$ is fixed by fitting the exchange energy with the Hartree-Fock data for noble gas atoms He-Rn and is found to be 0.0042. The functional is semiempirical in the sense that the parameter in it is obtained by fitting to known energies.

> Check that for $\rho(\boldsymbol{r}) \sim e^{-\alpha r}$, where $\alpha$ is a constant, the Becke functional gives $\epsilon_x(\boldsymbol{r}) = -\frac{1}{2r}$.

***Perdew's real space cutoff of the GEA hole***: Besides the condition on the asymptotic behaviour of the exchange energy per electron, one can also look to satisfy other conditions associated with the exchange energy. Two such properties are the normalization of the exchange hole and its non-negative value, expressed by Eqs. (4b) and (4c), respectively. However, as pointed out above, neither of these conditions is satisfied by the exchange hole corresponding to the GEA functional. When these conditions are enforced [51] on the exchange hole, the resulting energies are found to be superior to those given by the GEA. More importantly, the method shows the way to make progress towards more accurate functionals [53-55] by satisfying the exact conditions known about these functionals. Among these, the most widely used functional is the Perdew-Burke-Ernzerhof (PBE) functional [55]. The PBE exchange functional is

$$E_x^{PBE}[\rho] = \int \rho(\boldsymbol{r}) \epsilon_x^{LDA}(\rho) \left(1 + \frac{\mu s^2}{1+\frac{\mu s^2}{\kappa}}\right) d\boldsymbol{r} \quad , \tag{37}$$

where $\mu = 0.21951$ and $\kappa = 0.804$. Interestingly, a similar form for the exchange energy was earlier also proposed [56] by Becke with the coefficients $\mu$ and $\kappa$ determined by fitting the resulting exchange energy of atoms with their exact exchange energies. To include correlation energy in the functional above, the enhancement factor $F_x^{GGA} = \left(1 + \frac{\mu s^2}{1+\frac{\mu s^2}{\kappa}}\right)$ in Eq. (37) is replaced by a more general function $F_{xc}(s)$.

While performing self-consistent calculations with the GGA functionals, their functional derivative with respect to the density is required. These have been obtained [57,58] by applying the usual formula for functional derivative given in supplementary data of DFT-I.

## 2c. Going beyond the GGA

While GGA functionals give very good accuracy for exchange-correlations energies, they still do not satisfy many exact properties. As an example, the GGA functionals do not give potentials that are correct in the asymptotic regions of a system. GGA functionals are usually made more accurate in two ways. First, in addition to the density and gradient of the density, the functionals are made to depend on the KS kinetic energy density

$$\tau(\boldsymbol{r}) = \frac{1}{2} \sum_i |\vec{\nabla} \varphi_i(\boldsymbol{r})|^2 \quad . \tag{38}$$

Inclusion of $\tau(\mathbf{r})$ in the functional makes them dependent on the KS orbitals explicitly. These are known as **meta-GGA** (mGGA) functionals. Two examples of these are TPSS [59] and SCAN [60] functionals.

Second way of obtaining more accurate functionals is by mixing Hartree-Fock exchange energy calculated with KS orbitals with the exchange-correlation density functionals in some fixed ratio. This leads to

$$E_{xc}^{hybrid} = \alpha E_x^{HF}[\{\varphi_i\}] + (1-\alpha) E_{xc}^{DFT}[\rho] \quad \text{with} \quad 0 < \alpha < 1 \quad , \quad (39)$$

known as **hybrid functionals**. The density functional $E_{xc}^{DFT}[\rho]$ used in Eq. (121) can be the LDA or any of the other functionals discussed above. The idea of hybrid functionals was first proposed by Becke [61] who took $\alpha = \frac{1}{2}$ and $E_{xc}^{DFT}[\rho]$ as the LDA exchange-correlation functional. The proposal by Becke was based on the adiabatic-connection formula [62-65] for the exchange correlation energy. As is evident, hybrid functionals have explicit dependence on the KS orbitals. Two more examples of hybrid functionals are B3LYP [66,67], PBE0 [68]. Hybrid functionals differ from each other because of different mixture of Hartree-Fock exchange energy with other functionals and the correlation energy functionals they employ.

Another class of hybrid functionals is the **range-separated hybrid functionals** [69,70]. To construct these functionals, Coulomb interaction between electrons is split into its long-range and short-range components and the exchange-correlation energy is calculated as the sum of the Hartree-Fock exchange energy for the long-range and density functional exchange energy for the short-range part of the interaction and correlation energies. Rationale [70,71] for these functionals is that the approximate local density functionals for exchange-correlation energies are accurate for short range interactions while Hartree-Fock theory gives the correct long-range behaviour of the exchange-correlation potential.

Since meta-GGa and hybrid functionals have explicit dependence on KS orbitals, the corresponding KS potential cannot be directly obtained by taking functional derivative of these functionals with respect to the density. Furthermore, when total energy calculated using these functionals is minimized with respect to the orbitals, the resulting equation has orbital-dependent effective potential. Thus, the KS equation corresponding to these functionals is like Hartree-equation, i.e., with the effective being different for each orbital. This is referred to as the generalized Kohn-Sham equation [72,73]. If we wish to obtain a KS equation with single-multiplicative KS potential, it is done by using [74] the optimized potential method (OPM) [75-77] that will be discussed in section 3. As an example of generalized KS calculations in solids using meta-GGA, see reference [78].

Besides the functionals described above that are based on expanding the energy functional in terms of density gradient and kinetic energy density, there is another way of making exchange-correlation energy functionals more accurate and that is by making them self-interaction free. In the following we discuss this method in the context of local spin density approximation.

**2d. Self-interaction corrected local spin density approximation (SIC-LSD)**

In presenting some exact results in DFT above, we had discussed the exchange-correlation energy and the corresponding potential for a single-electron system. Since a majority of approximate energy functional do not give this correctly, it was suggested that these functionals can be made accurate [79,80] if they are corrected for the self-interaction error in them. This can be done by subtracting the self-interaction error for each orbital from the approximate functional. Thus, the self-interaction-corrected local-spin-density (LSD-SIC) exchange-correlation energy functional is [81]

$$E_{xc}^{LSDSIC} = E_{xc}^{LSD}[\rho_\uparrow, \rho_\downarrow] - \sum_i \left\{ E_{xc}^{LSD}[\rho_i, 0] + \frac{1}{2} \iint \frac{\rho_i(\boldsymbol{r})\rho_i(\boldsymbol{r}')}{|\boldsymbol{r} - \boldsymbol{r}'|} d\boldsymbol{r} d\boldsymbol{r}' \right\}, \qquad (40)$$

where $\rho_i(\boldsymbol{r}) = |\varphi_i(\boldsymbol{r})|^2$ is the single-electron density corresponding to the $i^{th}$ orbital. Notice that the functional above depends on the orbitals explicitly. Therefore, the corresponding equation for the orbitals has orbital-dependent potential in it and is a generalized Kohn-Sham equation. Results [80,81] obtained in the LSD-SIC formalism are much superior to the LDA results and remove some of the shortcomings of the latter. However, the correction is significant only if orbitals are localized and does not apply to extended orbitals. For solids, therefore, a transformation needs to be made to localized orbitals if the method is to be used there.

> Find the equation satisfied by the orbitals in LSD-SIC method by variational minimization of the energy with respect to the orbitals.

## 3. Obtaining exchange-correlation potential directly

### 3a. Modelling the potential

As we have seen in the discussion above, majority of exchange-correlation functionals do not give a potential that is correct in the asymptotic regions of a system. Although this does not affect the energy very much, it leads to the density in those regions and the highest occupied orbital eigenvalues to be highly inaccurate. As a result, properties that depend on these quantities - such as polarizabilities of atoms and molecules and band gaps of insulators - come out to be incorrect. To address this problem, another approach that has been employed is to model the exchange-correlation potential directly in terms of the density and its gradients. This is particularly useful in making the resulting ground-state densities, particularly their asymptotic behaviour, and the highest orbital eigenvalues closer to their exact counterparts. To model exchange-correlation potential directly, one is guided by the exact potential wherever it can be obtained (see sections 1b, 1d and 3a) and the functional forms already proposed for the exchange-correlation energy functionals in terms of the density or its derivatives. We discuss below two such potentials that have been used successfully, particularly when results require eigenvalues to be accurate, for example the band-gap problem.

*van Leeuwen-Baerends exchange potential*: Taking a cue from the Becke correction [46] to the LDA exchange functional, van-Leeuwen and Baerends (vLB) [82] introduced a similar correction to the LDA potential so that its asymptotic structure for finite systems is given correctly. The vLB potential, which by construction is the exchange potential, is

$$v_{x,m_s}^{vLB}(\boldsymbol{r}) = v_{x,m_s}^{LSD}(\boldsymbol{r}) - \beta\, \rho_{m_s}^{1/3}(\boldsymbol{r}) \frac{x_\sigma^2}{1 + 3\beta x_{m_s}\sinh^{-1}(x_{m_s})} \qquad , \qquad (41)$$

for electrons with spin component $m_s$ with constant $\beta = 0.05$. We recall that $x_{m_s} = \frac{|\nabla \rho_{m_s}|}{\rho_{m_s}^{4/3}}$. For spin unpolarized systems, the same form has been retained [83, 84] and the potential is given as

$$v_x^{vLB}(r) = v_x^{LDA}(r) - \beta\, \rho^{1/3}(r) \frac{x^2}{1 + 3\beta x \sinh^{-1}(x)} \quad , \quad (42)$$

with $\beta = 0.05$ and $x = \frac{|\nabla \rho|}{\rho^{4/3}}$.

> Show that for exponentially decaying densities, $v_x^{vLB}(r)$ has the correct asymptoyic beaviour for finite systems.

Using this potential in KS calculations has been shown to give [83] more accurate results for polarizabilities and hyperpolarizabilities of atoms. Furthermore, it leads [84-86] to accurate band gaps for a wide range of insulating and semiconducting materials.

***Becke-Johnson (modified-Slater) and modified Becke-Johnson potential***: The van-Leeuwen potential presented above depends on the density and its gradient, and may be called the GGA model potential because it is motivated by the Becke functional for the exchange energy. One can now go a step further, and in a manner similar to the development of functionals, include terms dependent on the kinetic energy density in the potential also, and develop mGGA model-potentials. One such potential was obtained by modifying the Slater potential, [see Eq. (8b)] by adding a correction term to it. The correction term was obtained [87,88] from the force field, calculated using Eq. (10), due to the Fermi hole corresponding [51] to the gradient expansion approximation (GEA). Becke-Johnson (BJ) [89] potential is precisely this potential and is given as

$$v_{x,m_s}^{BJ}(r) = v_{x,m_s}^{Slater}(r) + C_{\Delta V} \sqrt{\frac{\tau_{m_s}}{\rho_{m_s}}} \quad , \quad (43)$$

where $v_{x,m_s}^{Slater}(r)$ is the Slater potential and $\tau_{m_s}$ is the kinetic energy density, given by Eq. (38), for spin component $m_s$. The constant $C_{\Delta V} = \frac{1}{\pi}\sqrt{\frac{5}{6}}$ (There is a difference of a factor of 2 between the definition of kinetic energy density in ref. [89] and Eq. (38). This leads to the value of $C_{\Delta V}$ given in ref. [89] and that given here being different by a factor of $\sqrt{2}$).

> (i) Write the expression for $v_{x,m_s}^{Slater}(r)$ and $\tau_{m_s}$ in terms of the orbitals $\varphi_{i,m_s}$ for electrons of spin component $m_s$?
> (ii) Show that Eq. (43) gives the correct exchange potential for the homogeneous electron gas.

The potential $v_{x,m_s}^{Slater}(r)$ in Eq. (43) can be further approximated using the Becke-Roussel formula [90] and gives the potential

$$v_{x,m_s}^{BR}(r) = -\frac{1}{b_{m_s}(r)}\left[1 - e^{-x_{m_s}(r)} - \frac{1}{2}x_{m_s}(r)e^{-x_{m_s}(r)}\right] \quad , \quad (44)$$

where $x_{m_s}(r)$ is a parameter determined numerically by the density, its derivatives and the kinetic energy density at $r$, and $b_{m_s}(r)$ is calculated from $x_{m_s}(r)$. The Becke-Johnson potential is found [91] to give band gaps for semiconductors better than the LDA but not close to experiments. It was therefore further refined by Tran and Blaha [92] to obtain modified Beck-Johnson (mBJ) potential specific to calculation on solids. The potential is given as

$$v_{x,m_s}^{mBJ}(r) = v_{x,m_s}^{BR}(r) + (3c - 2)C_{\Delta V}\sqrt{\frac{2\tau_{m_s}}{\rho_{m_s}}} \quad . \quad (45)$$

Here $c$ is a constant calculated from the average value of $\frac{|\nabla \rho_{m_s}(r)|}{\rho_{m_s}(r)}$ over one cell of the solid and involves two parameters. For further details, we refer the reader to reference [92]. The mBJ potential leads [92,93] to accurate band gaps for a large number of insulating systems.

Next, we present a method that obtains the Kohn-Sham potential corresponding to orbital-dependent functionals, such the mGGA or LSDSIC functionals discussed above, in a mathematical rigorous manner.

### 3b. The optimized potential method (OPM)

To understand the motivation behind the optimized potential method (OPM), recall that the exchange potential in Hartree-Fock theory is a nonlocal potential. The question we ask now is if Hartree-Fock calculations can be performed with a local (multiplicative) exchange potential. The answer is in the affirmative and it gives rise to the optimized potential method (OPM) [75-77] for HF theory. The resulting local exchange potential obtained is then the exact exchange potential for a given system. Furthermore, as mentioned in section (2c) above, an additional advantage of the method is that it can be applied [74] to other orbital-dependent functionals to obtain the corresponding exchange-correlation potential for functionals, such as the LSDSIC [94] or the mGGA functionals [74], which depend on orbitals explicitly. In the following we give a brief presentation of how the OPM is carried out for a general energy functional $E_{xc}[\varphi_i]$ in which the exchange-correlation energy is expressed in terms of the orbitals. Consider a set of occupied orbitals generated by solving the equation

$$\left(-\frac{1}{2}\nabla^2 + V(r)\right)\varphi_i(r) = \epsilon_i \varphi_i(r) \quad , \quad (46)$$

where the potential $V(r)$ is to be determined by the condition that the orbitals obtained from Eq. (46) minimize the energy functional $E[\varphi_i]$, which is the sum of the kinetic energy, the external energy, the Hartree energy and the approximate exchange-correlation energy. This is done by minimizing $E[\varphi_i]$ with respect to variations in the potential $V(r)$. This amounts to setting the functional derivative

$$\frac{\delta E[\varphi_i]}{\delta V(r)} = \sum_i \int \frac{\delta E[\varphi_i]}{\delta \varphi_i(r')} \frac{\delta \varphi_i(r')}{\delta V(r)} dr' = 0 \quad . \quad (47)$$

This in turn leads to an integral equation for $V(r)$. Using that equation for updating $V(r)$, Eq. (46) is solved self-consistently and leads to the potential and orbitals that give the minimum value for the functional $E[\varphi_i]$. This then is the ground-state energy for the system given by the exchange-correlation energy functional employed; $V(r)$ and $\{\varphi_i(r)\}$ are the corresponding Kohn-Sham potential and orbitals. The exchange-correlation potential for this functional is then obtained by subtracting the external and the Hartree potentials from $V(r)$.

The OPM was first employed [76] to get a local potential to minimize the Hartree-Fock energy of atoms and therefore gave the exact KS exchange potential [95] and orbitals for these systems. Since then, it has been applied to molecules [96] and solids taking only exchange [97-100] into account and also employing exchange mixed with correlation [100-102], and the mGGA functionals [74,78]. With this we conclude our discussion of the optimized potential method. For an extensive review of the OPM and its application to a variety of orbital-dependent functionals, we refer the reader to reference [103].

**3c. Application to calculation of band gaps of different solids**

To provide a perspective on the results obtained when the exchange-correlation potential is calculated through various methods described above. Thus, the potential is either modelled in terms of the density and its gradient or calculated from the orbitals through the HS potential of Eq. (10) calculated within the exchange-only or by employing the OPM. We focus on the band gap because it can vary by large amount when calculated using different methods - other quantities of interest, such as the lattice constants or the bulk moduli come out to be nearly the same [84, 86] in different potentials.

Shown in Table II are the band-gaps of various materials obtained by applying the potentials described above. HS-EX refers to calculations [84] done by applying the HS exchange potential, vLB implies results of applying [84, 85] the van Leeuwen-Baerends exchange potential together with the LDA correlation potential, under the column MBJLDA are given the results [92,104,105] obtained from modified Becke-Johnson exchange potential combined with the LDA correlation potential and the EXX gives the results [98-101] of employing the exact exchange potential using the OPM. For comparison the LDA results and those obtained [106-109] by the use of the GW method are also given. All these numbers are compared with the experimental results [110-114]. For further discussion and more calculations on solids using different kinds of functionals and optimized potential method, some more studies we refer the reader to are [115-120]

It is clear from the results presented in the Table that or the majority of materials studied, the exchange potentials of HS and the OPM, as well as the model potential lead to results that compare well with the experimental results. Interestingly, these results are obtained with numerical effort which is much less compared to the computationally demanding GW method, leading however to numbers of comparable accuracy.

Calculation of band-gaps employing Kohn-Sham method is a challenging and open problem for further investigations. This is mainly because of the derivative discontinuity in the exchange-correlation potential. In this connection, we refer the reader to a recent study [121] of the KS system for Si and NaCl. The KS systems for these were constructed by inverting their accurate densities obtained [122] by Quanum Monte-Carlo method.

**Table II:** Band gaps of several materials calculated by applying the LDA, HS-EX [84], vLB [84,85], MBJLDA [92,104,105], EXX [98-101] and GW [106-109] are compared with experiments [110-114]. Result for applying MBJPBE functional for BP [118] is also shown for completeness.

| System | Band-gaps [eV] |
|---|---|

|       | LDA  | HS-EX [84] | vLB [85,86] | MBJLDA [92] | EXX        | GW [106, 107]          | Expt.        |
|-------|------|------------|-------------|-------------|------------|------------------------|--------------|
| Ne    | 11.39| 22.07      | 23.64       | 22.72       | 14.15 [100]| 22.1 [107]             | 20.75 [110]  |
| Ar    | 8.09 | 11.29      | 12.76       | 13.91       | 9.61 [100] | 14.9 [107]             | 14.32 [110]  |
| Kr    | 6.76 | 9.10       | 10.91       | 10.83       | 7.87 [100] | -                      | 11.40 [110]  |
| Xe    | 5.56 | 6.63       | 8.61        | 8.52        | 6.69 [100] | -                      | 9.15 [110]   |
| C     | 2.70 | 5.47       | 5.18        | 4.93        | 5.12 [98]  | 6.18 [107]             | 5.48 [111]   |
| Si    | 0.49 | 1.24       | 1.21        | 1.17        | 1.93 [98]  | 1.14 [106]; 1.41 [107] | 1.17 [111]   |
| MgO   | 4.94 | 6.23       | 6.94        | 7.17        | 7.77 [99]  | 9.16 [107]             | 7.78 [111]   |
| CaO   | 3.36 | 7.29       | 7.15        | –           | 7.72 [99]  | -                      | 7.09 [111]   |
| LiF   | 8.94 | 9.52       | 12.61       | 12.94       | –          | 15.9 [107]             | 13.60 [112]  |
| LiCl  | 6.06 | 6.50       | 7.84        | 8.64        | –          | -                      | 9.40 [111]   |
| AlN   | 2.44 | 5.05       | 5.13        | 5.55        | 5.03 [101] | 4.90 [108]             | 5.11 [113]   |
| AlP   | 1.16 | 2.53       | 2.75        | 2.32        | –          | 2.90 [107]             | 2.51 [114]   |
| BP    | 1.51 | 2.22       | 2.09        | 1.95 [104]  | –          | -                      | 2.00 [114]   |
| 3C-SiC| 1.38 | 2.88       | 2.58        | 2.52 [105]  | –          | 2.76 [109]             | 2.42 [113]   |

## 4. Orbital free DFT

In the presentation above, we have mainly focused on calculations within the realm of Kohn-Sham theory in which the exchange-correlation potential corresponding to an exchange-correlation energy functional is obtained as its functional derivative with respect to the density. Exceptions to this were the mGGA and LSDSIC functionals where the potential was orbital dependent. Nonetheless, calculations were always done in terms of orbitals. In this section, we discuss the development of calculation schemes purely in terms of the density.

In DFT – I we had introduced Thomas-Fermi (TF) theory in which calculations are done entirely in terms of the density, albeit approximately. After the Hohenberg-Kohn theorem was proved, it became clear that TF theory is an approximation to the exact DFT. Thus, it is the earliest form of DFT that employed only the density and was therefore orbital free. Such theories are given the name orbital-free density functional theory (OF-DFT) and in principle offer better scaling with the number of electrons. Euler equation for the density in TF theory has already been presented in DFT-I. To go beyond TF theory, one must develop noninteracting kinetic energy density-functionals that are accurate - just like it is done for the exchange-correlation energy functionals using the derivatives of the density - to improve on the LDA for these energies. As discussed in DFT-I, the earliest modification of this kind to the TF kinetic energy is the von-Weizsäcker correction [123] involving the gradient of the density. Gradient expansion approximation for kinetic energy up to fourth order in the gradient of the density has also been developed [124] and studied [125, 126]. Sixth order formula for the kinetic energy has also been derived [127]. References [128-130] give examples of some recent kinetic energy functionals and their applications. In this paper, we present a different form of the equation for the density. The equation is like the Schrödinger equation for the square-root of the ground-state density and is therefore easy to implement using existing KS codes.

The idea behind obtaining the equation for the square-root of the density is as follows. First write the kinetic energy as the sum of the von-Weizsäcker energy functional $T_{vW}[\rho]$ and what is called [131] the Pauli kinetic energy $T_\theta[\rho]$ (this is the difference between the non-interacting kinetic energy and the von-Weizsäcker energy). Thus,

$$T_s[\rho] = \frac{1}{8}\frac{|\nabla\rho(\boldsymbol{r})|^2}{\rho(\boldsymbol{r})} + T_\theta[\rho]$$

$$= \frac{1}{2}\left|\nabla\sqrt{\rho(\boldsymbol{r})}\right|^2 + T_\theta[\rho] \qquad (48)$$

With the noninteracting kinetic energy written in the form above, minimizing the energy functional with respect to $\sqrt{\rho(\boldsymbol{r})}$ with the constraint that the total number of electrons is fixed leads to the equation

$$\left[-\frac{1}{2}\nabla^2 + v_\theta(\boldsymbol{r}) + v_{ext}(\boldsymbol{r}) + v_H(\boldsymbol{r}) + v_{xc}(\boldsymbol{r})\right]\sqrt{\rho(\boldsymbol{r})} = \mu\sqrt{\rho(\boldsymbol{r})} \quad , \quad (49)$$

where all the symbols have their standard meaning and there is a new term

$$v_\theta(\boldsymbol{r}) = \frac{\delta T_\theta[\rho]}{\delta \rho(\boldsymbol{r})} \quad , \qquad (50)$$

which is known [131,132] as the Pauli potential. The equation above has also been obtained in other ways [133-135]. As example of applications of the equation above, we refer the reader to references [134,136].

> What is the Pauli kinetic energy and the corresponding Pauli potential for the ground-state of a single-orbital system like He?

Listed below are some of the advantages of OF-DFT:

I. Reduced number of degrees of freedom.
II. Without any orbital dependence, the complication and cost associated with orbital manipulation, including orbital orthonormalization and orbital localization (for linear scaling implementations) are avoided.
III. For metals, the need for Brillouin-zone (k-point) sampling of the wavefunction is completely eliminated.
IV. The utilization of the fast Fourier transformation in solving the orbital-free model is essentially linear scaling with respect to the system size.

Before we leave this section, we mention two major challenges that implementation of OF-DFT faces. First one is of course the development [137] of accurate orbital free kinetic energy density functionals (OF-KEDF) and their effective implementation [138]. The second significant challenge for application of OF-DFT is to construct accurate and transferable local pseudopotentials for each element [139-141]. This limits the employability of nonlocal pseudopotentials in OF-DFT. Recently, nonlocal pseudopotential (NLPP) approach proposed by Xu et al [142] has shown great promise than local pseudopotential methods. To overcome the over reliance on all-electron potential, the NLPP adds a nonlocal energy term with a set of angular-momentum-dependent energies unlike orbital-based approaches that ignore nonlocal energy terms [143].

Given that all above points are satisfied, in the future it may be expected that OF-DFT will allow us to simulate much larger systems than Kohn-Sham DFT although one may have to sacrifice some accuracy in doing so.

## 5. Concluding Remarks

In this article, we have discussed many fundamental aspects of DFT that lead to a better understanding of the theory and help in applying it more effectively across a range of systems. Thus, we started with a description of some exact results in DFT, focusing mainly on the exchange-correlation energy functional and the corresponding potential. In the process we also introduced the inverse Kohn-Sham procedure and applied it to some small systems to obtain their exact exchange-correlation potential. Through this, we also demonstrated the ionization-potential theorem of DFT. In our discussion, we used the physical description of the exchange-correlation potential given by Quantal DFT. This explains the mathematical prescription of exchange-correlation potential in terms of the Coulombic and kinetic-energy related force fields arising due to many-body effects.

An important aspect of the exact DFT description arises when it is generalized to systems with fractional number of electrons. We have discussed this and the resulting derivative-discontinuity of the exchange-correlation potential in detail. This generalization and the derivative discontinuity of the exchange-correlation potential give very simple results but have subtleties that we hope are clarified by our discussion. A significant consequence of the derivative discontinuity of the potential is that the fundamental gap of semiconductors and insulators is underestimated even in the exact Kohn-Sham theory. We have discussed this after the section presenting DFT for fractional number of electrons.

Having described the exact properties of the exchange-correlation functionals and the corresponding potentials, we next discussed how the functionals are made more accurate going beyond the LDA by incorporating these properties while constructing new functionals. This has given rise to self-interaction-corrected LDA, GGA and meta-GGA functionals. Another approach to designing accurate functionals we described is that of mixing exact exchange energy calculated in terms of KS orbitals with approximate density-based functionals, which has led to hybrid functionals and range-separated hybrid functional. The mixing is done on the basis of adiabatic connection formula for the exchange-correlation energy functional.

A different approach that is also popular for performing DFT calculations is to model the exchange-correlation potential directly bypassing taking the functional derivative of a functional.

The potentials are constructed so that they satisfy certain exact properties to be satisfied by the exchange-correlation potential and to give certain properties – we focused on the band-gap of semiconductors – of the systems being studied accurately. We discussed two such model potentials and results they gave for the band-gaps. While on the subject of modelling exchange-correlation potentials, we also described the optimized potential method for constructing exact Kohn-Sham potential for orbital-dependent functionals.

The original motivation for DFT was to be able to perform electronic structure calculations treating ground-state density as the basic variable. However, lack of accurate kinetic-energy density functionals led to KS formulation of DFT and that has been the mainstay of DFT calculations since its inception. Nonetheless, slowly efforts are being made to develop orbital-free DFT that facilitate electronic structure calculations of very large systems. Our final discussion in section 4 describes the development of these methods along with the main challenges in its implementation.

Density-functional theory is a vast subject, and it is not possible to cover all its aspects in the two articles (DFT-I and the present one) that we have written. The material presented thus reflects both our subjectivity and limitations and we admit that with humility. At the same time, we hope that the articles have given the readers wide enough view of the subject. We would like to end this article by pointing out two important topics that have not been discussed and are equally important. One is time-dependent DFT (TDDFT) and the other is use of machine learning in DFT.

TDDFT is based on Runge-Gross theorem [144] which states that if all systems start from the same initial wavefunction and the external potential can be expanded in a Taylor series in time around the initial time, then there is one-to-one mapping between the potential and the resulting density. The theory had earlier been formulated [145] for potential periodic in time using Floquet [143,146,147] approach to solving time-dependent Schrödinger equation for such potentials, and by using [148,149] minimization principle for energy functional in adiabatically switched-on potentials employing hydrodynamic formulation of time-dependent Schrödinger equation. TDDFT has been applied successfully to study a range of time-dependent phenomena. A particular application of the theory has been to obtain excitation energies – both involving single particle excitations and collective excitations – of many-electron systems using linear response. We refer the reader to reference [150] for learning more about fundamentals and applications of TDDFT. A recent review [151] discusses calculations [152,153] of collective oscillations using orbital free TDDFT.

On the computational aspect of applying DFT, there have been several developments in providing numerically efficient and accurate solution to KS-DFT. Artificial intelligence (AI) and machine learning (ML) based approaches are such future looking methodologies employed to solve KS-DFT to obtain electronic-structure of atoms, molecules, and solids [154-157]. Yao and Parkhill [158] have developed a very interesting ML approach, where they used 1D convolutional neural network to fit the kinetic energy as a function of the density projected onto bond directions. Brockherde et al. [159], on the other hand, have shown the possibility of bypassing KS equation using AI/ML in predicting electronic structures with lower computational cost. While Nagai et al used neural network-based ML framework to accurately solve Kohn-Sham equation for atoms [160].

To sum up, use of AI/ML in solving electronic-structure problem shows the way to the future as models trained on certain chemical elements or phases show applicability to materials with the same constituent and compositions. For example, Nagai et al. [160, 161] applied AI/ML to improve the accuracy of the XC functional in the KS equation. The problem, however, remains challenging due to unavailability of accurate and large-enough database and new functional training and development for any practical purposes will open the gateway to its future applications.

**Acknowledgement**: The authors thank the editors for invitation to write the article presented here. Work at Ames Laboratory was supported by the US Department of Energy (DOE) Office of Science, Basic Energy Sciences, Materials Science & Engineering Division. Ames Laboratory is operated by Iowa State University for the US DOE under contract DE-AC02-07CH11358.

**Author contributions**: A.K.: data curation, investigation, formal analysis and writing—original draft, review and editing. P.S.: data curation, investigation, formal analysis and writing—original draft, review and editing. M.K.H.: conceptualization, supervision, data curation, investigation, formal analysis, writing—original draft, review and editing.